\documentclass[mathleft
]{an}
\usepackage{graphicx}
\usepackage{times}
\usepackage{hyperref}
\begin{document}

\Pagespan{280}{}
\Yearpublication{2012}%
\Yearsubmission{2012}%
\Month{2}%
\Volume{333}%
\Issue{3}%
 \DOI{10.1002/asna.201211656}%

\title{
        Career situation of female astronomers in Germany}

\author{J. Fohlmeister\inst{1}\fnmsep\thanks{Corresponding author:
  {janine@ari.uni-heidelberg.de}}
\and  Ch. Helling\inst{2}
}
\titlerunning{AstroFrauenNetzwerk Survey}
\authorrunning{J. Fohlmeister \& Ch. Helling}
\institute{
Astronomisches Rechen-Institut, Zentrum f\"ur Astronomie der Universit\"at Heidelberg, M\"onchhofstra\ss e 12-14, D-69120 Heidelberg, Germany 
\and 
SUPA, School of Physics and Astronomy, University of St. Andrews, North Haugh, St Andrews KY16 9SS, UK
}

\received{2012 Feb 10}
\accepted{2012 Mar 2}
\publonline{2012 Apr 5}


\abstract{%
We survey the job situation of women in astronomy in Germany and of
German women abroad and review indicators for their career
development. Our sample includes women astronomers from all academic
levels from doctoral students to professors, as well as female
astronomers who have left the field. 
We find that
networking and human support are among the most important factors for
success. Experience shows that students should carefully choose
their supervisor and collect practical knowledge abroad. We reflect
the private situation of female German astronomers and find that
prejudices are abundant, and are perceived as discriminating. We identify 
reasons why women are more likely than men
to quit astronomy after they obtain their PhD degree.
 We give
recommendations to young students on what to pay attention to in order
to be on the successful path in astronomy.}

\maketitle

\section{Introduction}

Although astronomy is regarded as more attractive to women compared
to physics, the number of women who strive for a career
in astrophysics is reduced significantly with each career step,
resulting in a high degree of underrepresentation at professorial
level. Woman in general do not have a strong presence in science 
and technology, with the proportion of woman varying widely within Europe.
In Germany, women are known to have less interest than men in natural 
science than in other EU countries (e.g. \cite{ainu}). The percentage of 
female students in physics (that provide the pool for future astronomers) 
at German universities is of ~20\,\% since years.

In the past and in the present, women astronomers had and have
the choice of blending in with their environment or just accept to be
different. Many of us were wondering for a good fraction of our career
if our experiences, for instance in feeling as the odd-one-out during
an opening address like {``Good morning
Sirs''} or by aiming for quality rather than self-assertion, are
singularities.  Should we just learn to adjust, and become less
thin-skinned when being confronted with degrading comments and
prejudice because this is what happens when climbing the hierarchy?
Do we have to work harder in order to be promoted?  Are we really
confronted with discrimination or just too sensitive when meeting a
bad-tempered person -- one out of so many brilliantly intelligent,
helpful and supportive colleagues of both genders?

Factors that affect the hiring of female faculty candidates and carreer barriers in natural sciences 
have been mainly studied in the US (e.g. \cite{greene}).
To quantify the status of women in astronomy in Germany and of German
women abroad the AstroFrauenNetzwerk (AFN) in Germany has
conducted two surveys over the last two years. The AstroFrauenNetzwerk
was founded in 2007 as an independent network for female astronomers
who hold a PhD in Germany or have obtained their PhD in
Germany. Its aim is to connect women astronomers, to collect and
share relevant information and to help its fellow members in and from
a country where women are traditionally underrepresented on
managerial levels not to mention the professorial level. Meanwhile the
AFN has 73 regular members of various nationalities working in or
originating from Germany plus 3 associated international members. Two of the
founding members of the AFN now hold permanent posts outside Germany.

In a first study in 2010, a sociologist worked with the
AstroFrauenNetzwerk evaluating the gender distribution in astronomy
in Germany.  Data were collected from German astronomy institutes\footnote
{Some German institutes questioned our right to collect these data.} and
their web pages, and the annual reports of the Astronomische
Gesellschaft.\footnote{
\url{www.astro.rug.nl/~kamp/AFNmeeting2010.html}}  The
percentage of women in German astronomy was found to be about 20\,\%
among students, which is comparable to the proportion of physics in
general. There is an increased number of female PhD students compared
to graduates in astronomy, mainly due to the recruitment of
international students. At postdoc level the percentage of women drops
to 17.2\,\% and becomes fractional at professorial level with 3.8\,\%.

In a second step, we used our own, independent pool of women
astronomers to produce statistics regarding the status of women in
astronomy in Germany and of German women abroad and indicators for the
career development of these women. Here we report the latest results
of a comprehensive survey conducted in autumn 2011.

Before presenting our results, we are sketching the bigger picture of
German women in natural sciences and astronomy as it emerges from
recent statistics of the German Research Foundation (DFG), the
International Astronomical Union (IAU), the Max Planck Society (MPG), 
the German Physics Society
(DPG), the American Institute of Physics, and the University of
Heidelberg.

According to the DFG statistics that is based on material from
proposal evaluations, between 2005 and 2008 only 6.8\,\% DFG proposals
for individual grants in astrophysics and astronomy were submitted by
women in Germany. The statistics for the year 2010 published by the
DFG show that for all natural sciences the funding rates for all
branches was somewhat lower for women (42.9\,\%) than for men
(45.8\,\%). The principle investigator-ship (PI)  of women in DFG
research units (Forschergruppen) proposals was 8\,\% in natural sciences
and 8.7\,\% for PI-ship in the framework of the Excellence Initiative of
the German universities.\footnote{For more details and additional data see
\url{www.dfg.de/download/pdf/foerderung/.}} The numbers
from the DFG statistics refer to submitted applications only. The
question here is: Are these numbers representative for women in the 
respective scientific level? 

Regarding the presence of women in the International Astronomical
Union (IAU), Germany holds one of the last positions (31 out of 36
national members, \cite{cesarsky}) with 9.1\,\% women members in 2010.\footnote{See 
\url{www.iau.org/administration/membership/} for the geographical distribution.} Individual
Membership in the IAU is open to scientists holding a PhD and who are
``eligible for election as individual members`` admitted by a national
member (in Germany: Rat Deutscher Sternwarten, in UK: Royal
Astronomical Society, etc.). This means to become a member of the IAU
the person has to have a certain visibility to be promoted. If women
have difficulties with recognition or showing initiative they will be
underrepresented in the IAU compared to the female astronomers
population in Germany.

The local organizing committee of the annual meeting of the
Astronomische Gesellschaft in Heidelberg in September 2011 counted
20\,\% women among the 430 participants. Out of all female participants
approximately half currently work in Germany. 32\,\% (6 in numbers) of
the invited plenary talks were given by women, including a high fraction of international female astronomers. 

The Max Planck Society, one of Germany's leading research organizations, 
operates nine research institutions in Germany that carry out research in astronomy and astrophysics.
These institutions are lead by a total of 32 directors, not one of them is female.\footnote 
{http://www.mpg.de/institutes}

The questions that follow are: where have the female astronomers from Germany gone before reaching 
top level? 
Are they (consciously or unconsciously) less promoted after PhD and/or do they
prefer to leave either the country or astronomy? 

In a survey initiated by the Equal Opportunities Study Group of the
German Physics Society (DPG) in 2002 among 1500 male and female
physicists and astrophysicists in Germany, men with children were
found to receive the greatest opportunities in their profession,
followed by men without children and women without children. Women
with children were found off the beaten track. The latter suffering
reservations on the part of employers (who are on the majority male)
fearing caring for a family will affect their performance at
work. Moreover, the DPG study found that women were more rarely
in executive positions and had lower income than their male colleagues (\cite{koene}).

A similar but world-wide survey of 15\,000 physicists by the American
Institute of Physics found that women are less likely than men 
to report access to various resources and opportunities that would be 
helpful in advancing a scientific career (international opportunities, invitations to
speak, supervisory experiences, serving on committees that have
influence, editor of journal, advising students) and are more often
assigned to less challenging work by their employer when they became a
parent (\cite{cesarsky}, see also www.aip.org/statistics).

A recent study at the Department of Physics and Astronomy in
Heidelberg, which is the largest of its kind in Germany, showed that
women more often fear failure than men, although no difference in
their qualification and commitment for a physics career was
found. Reasons are a mixture of missing support
and subsequent self-underestimation (\cite{fohlmeister}), 
which is alarming given the demographic challenges Germany is facing in the 
coming years.

Here we focus on the career status of German female astronomers. Our
sample includes women astronomers from all academic levels from
doctoral students to professors, working in Germany and abroad, as
well as female astronomers who have left the field. In this study we
examine the situation of female astronomers and identify important
criteria for a career in astrophysics. We investigate private and
professional impact factors as well as reasons to work or not to
work abroad. We also show the comments and cite three example
situations female astronomers are still confronted with, and hope to
increase the awareness that comments and behavior like these
are discriminating and socially retarded.

\section{Survey description}
The survey presented here was conducted by collecting data via an
online-questionnaire.  This questionnaire was designed in preparation
of the annual meeting of the AstroFrauenNetzwerk (AFN) in September
2011. The invitation to participate in the survey was sent to all
members of the AFN. The members were also asked to forward the
invitation to participate to all women astronomers in their respective
institutes and to female German-speaking colleagues world-wide. To
participate in the survey questions had to be answered online
anonymously. Altogether 61 women filled in the questionnaire.  No
reliable statistics are available on how many German-speaking female
astronomers work in Germany and abroad. Comparing the number of
participants in this study (61) to the number of all AFN-members (73),
we estimate that the participation completeness of this survey is at least 80\,\% . We
note that not all AFN members filled the questionnaire and not all
women who filled the questionnaire are AFN members.

\subsection{Target group for survey}

Our data define a representative sample of the German astronomical
community. It includes female astronomers from all academic levels
ranging from doctoral students to professors, as well as female
astronomers who have left the field and work outside astronomy.

The age distribution of the participants is Gaussian, with a peak
around 31--35 years. Due to the composition of the AFN (women can
become members after obtaining their PhD), the fraction of postdocs is
largest within this survey. It also includes a fair fraction of female
junior research group leaders (5 in numbers) and professors (4 in
numbers). In this sample the category ``professor'' applies to all
professorial titles, e.g. full professorships as well as German
``apl. Professor'' (adjunct professor). Their small total numbers among the participants in this
survey reflects the underrepresentation of women in the German
academic system. Currently only two full university and three associate 
astronomy professorships are filled with women in
Germany. There exist different forms of junior research group
positions in astronomy. Only DFG-funded research groups in the Emmy
Noether program list the number of female research group leaders in astronomy
(currently 5 out of 13 in total).

Out of all woman who participated in the survey, more than two third
(70\,\%) grew up in Germany, 27\,\% in other countries in Europe and 3 \%
overseas.

\section{Survey results}

\subsection{Current and desired work place}
73\,\% of the survey participants worked in Germany when they filled in
the survey, 20\,\% in other European countries, several (7\,\%) outside
Europe. We were interested if Germany is seen as an attractive work
place to perform astronomical research. Therefore all participants in
the survey who are not located in Germany were asked, if they would
like to work in Germany and why. About one third (27\,\%) answered with ``yes''
and gave private reasons for their decision such as cultural binding
or that they wish to be close to family and friends. About 20\,\%
answered with ``no''; mainly due to work related reasons, stating that
they prefer the better working conditions, having more scientific
freedom, working in countries with better research politics, or the
teaching approach in their respective country of work.  Some of the
women working outside Germany dislike German attitude (``Germany is too
German", in USA ``weniger Gemecker" (less nagging)).

The duration of the current contracts of all survey participants shows
a flat distribution ranging from 1 to 5 years with a small peak around
3 years (25\,\%). Only 14\,\% of the women who responded hold a permanent
position (8 in numbers).  Nevertheless, 32\,\% are satisfied with their
current job situation and 35\,\% rank their job as ``short transitional
phase", 40\,\% as stable phase and 25\,\% as ``long-term perspective".

\subsubsection{Finding a job}

Our results show that the way of finding a job in astronomy is
two-fold in Germany, either by the path
of job advertisements or applying to soft money, or else by
personal contacts.
  
This is reflected in the numbers:
German-speaking female astronomers find their job mainly due to
personal contacts (46\,\%) and/or the classical advertisement on the
internet (40\,\%). About 20\,\% have changed their position within the
institute. The percentage of women who were personally asked to come
for the job is the same as the percentage of those who raised their
own money (both 18\,\%).

\begin{figure}[t]
\includegraphics[width=8.3cm,angle=0,clip]{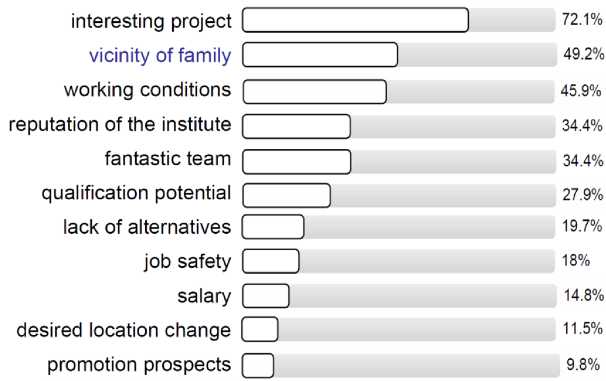}
\caption{``Why did you decide for your current position?"
Percentage of women who selected above factors (multiple answers were possible),
sorted by frequency of mention. The only not work-related reason of choice, ``vicinity of family", is ranked second highest.}
\label{fig1}
\end{figure}

We also wanted to identify what the most important criteria on which
women base their decision for a job are. The survey participants
decided for a position mainly based on how interesting a project
appeared (Fig.~\ref{fig1}). The second most important factor is vicinity of
family (this includes spouses and all kind of relatives like parents
or children), followed by working conditions or the reputation of the
institute. Job safety, the salary, a desired change of location or the
offered prospect of promotion are among the less decisive factors.

\subsubsection{Satisfaction and contentment}

\begin{figure}
\includegraphics[width=83mm]{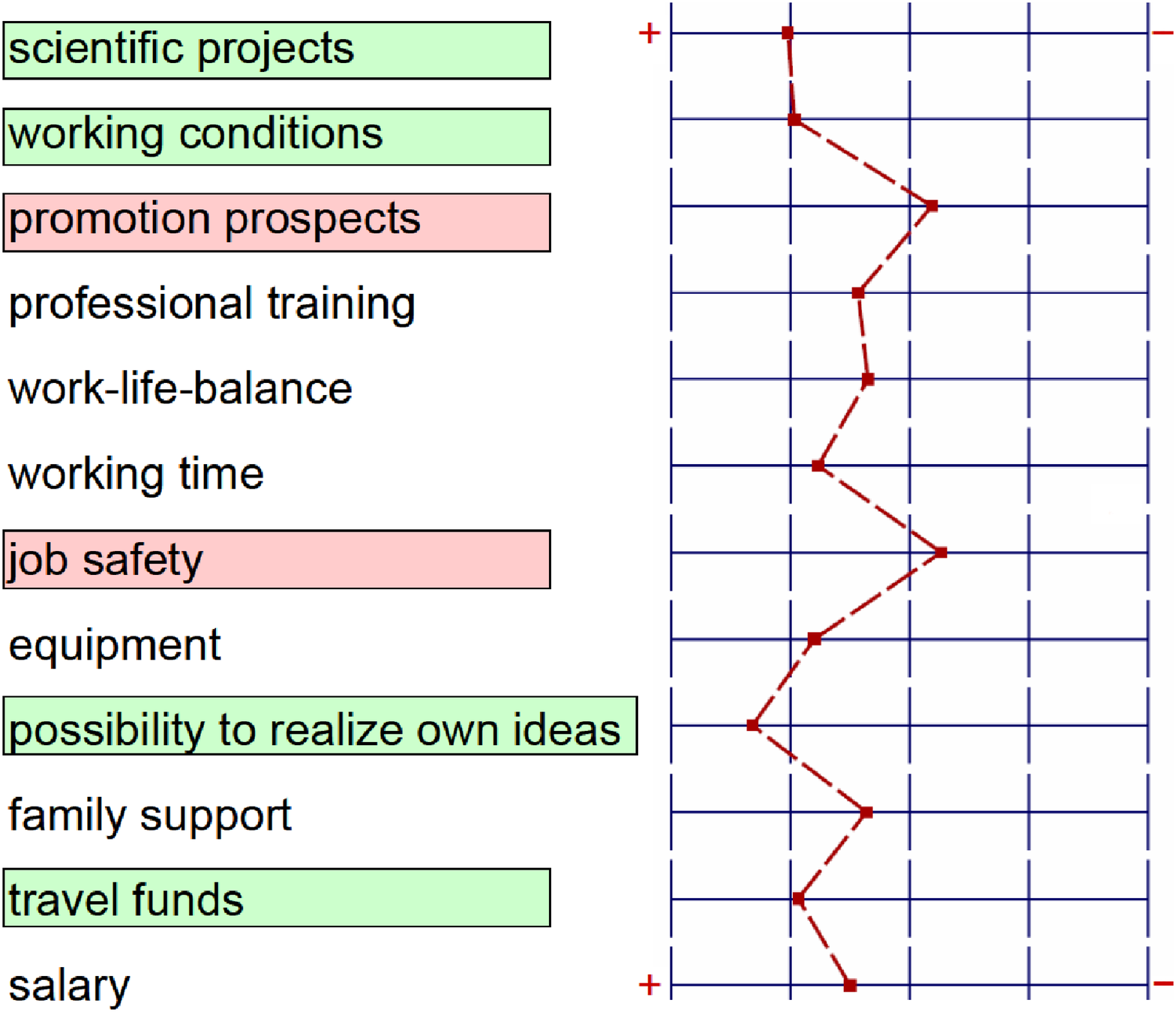}
\caption{(online colour at: www.an-journal.org) ``Are you happy with your ...?" 
Left is ``very happy", right ``not at all".  Average contentment of all survey participants with different job aspects.}
\label{fig2}
\end{figure}

We  asked about the satisfaction and how content the women were
with different job aspects (see Fig.~\ref{fig2}). Most female astronomers
are happy about the possibility of carrying out their own ideas, the
availability/amount of travel funds, their scientific projects and
working conditions.  These match the above selection criteria in
Fig.~\ref{fig1}.

On the other hand, female astronomers are less satisfied with their
promotion prospects and job-safety.  We note that these two points
were not decisive for choosing the job. Other criteria like
professional training, work life-balance, working time, equipment and
family support, and salary are neither ranked very high nor very low.

42\,\% of the female astronomers in our sample state that they want to
continue research in astronomy in the long term, 55\,\% if
possible. Only 3\,\% say they ``don't need to". One woman stated: ``Under
current circumstances my answer is: yes, if possible, but not at any
cost. With just a slightly better perspective for a permanent position
I would say: yes, absolutely- dream job. The result above might also
be biased in the sense that woman who do not want to stay in astronomy
did not participate in the survey.

\subsubsection{Salaries}

The average monthly net income summarized here can be treated as an
orientation on what young females 
should expect 
as a minimum in astronomy in Germany. Global
surveys suggest a note of caution as they show that men have generally
higher incomes. We note that the average numbers from our sample do
not reflect the individual salaries, which strongly depend on the
source of funding, the employer and regional differences. For PhD
students the average monthly net income in the sample is 1255 EURO, for
postdocs 2080 EURO and for staff astronomers 3210 EURO. We note that the
salaries of female astronomers who work outside academia show an
extremely broad range. This also holds for female professors,
depending on successful negotiation.

\subsection{Private life}

About 77\,\% of the female astronomers in our sample have a partner. Out
of these, every second woman is engaged with an astronomer or
physicist. The spouses of the remaining participants work outside
sciences (44\,\% out of 77\,\%) or in the minority (6\,\% out of 77\,\%) doing
other science than physics or astronomy. So, if female astronomers
choose a partner, every second woman goes for a ``colleague". Other
scientists than astronomers or physicists are the exception. In
90\,\% of all cases in our sample the partner lives in the same city,
10\,\% have long-distance relationships.

\subsubsection{Motherhood}
Being mother applies only to one third of our sample, nevertheless
90\,\% would like to have at least one child in the future. Reasons for
not having children yet are a mix of job-related and private
reasons. 60\,\% of these women refer to the uncertain job
situation. Around one third states the lack of right partner or that
they do not see how to arrange work and family, or that the work-load
is too high. About 10\,\% do not want to have children.  Those who
became mothers feel very restricted in mobility and point out that it
is harder to combine job and family. However, mothers feel generally
more balanced and motivated.

\subsection{Career development}

We asked how many stays abroad for more than 3 months have been
carried out. Despite the career importance of working abroad, 29\,\% of
the female astronomers in the sample never had a stay abroad so
far. All female staff, professors and astronomers with permanent
positions in this survey had at least one stay abroad for more than 3
months.

\begin{figure}[t]
\includegraphics[width=8.3cm,angle=0,clip]{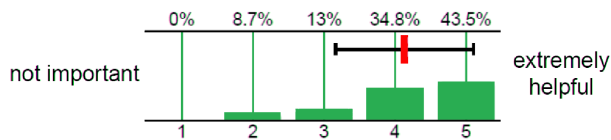}
\caption{(online colour at: www.an-journal.org) Left is ``Mentor was not important", right "extremely helpful for the career".  
Percentage of answers (on top) and mean value (red bar) of all female astronomers who had a mentor.}
\label{fig3}
\end{figure}

More than 40\,\% of the female astronomers in our sample have been
supported by a mentor, mainly male mentors (70\,\%) and state that
having a mentor was very helpful for their career (Fig.~\ref{fig3}). Note,
that these numbers do not refer to a professional mentoring scheme.

\begin{figure}[t]
\includegraphics[width=8.3cm,angle=0,clip]{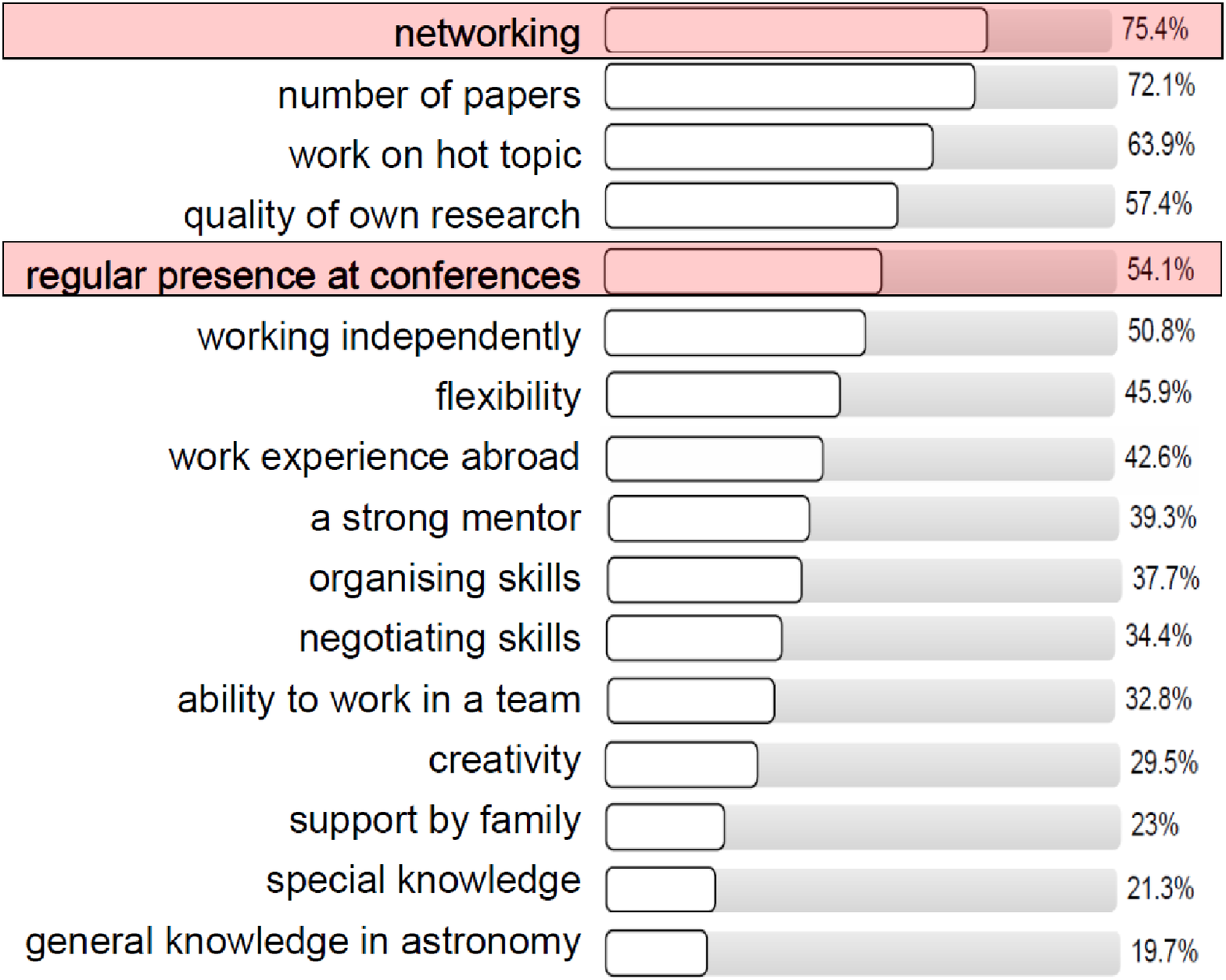}
\caption{(online colour at: www.an-journal.org) ``What are the most important
factors for a career in astronomy?" Percentage of women who selected above
factors (multiple answers were possible), sorted by frequency of mention.
``Networking" and ``regular presence at conferences" are the favorites in the subgroup of female professors.}
\label{fig4}
\end{figure}

It was consensus (see Fig.~\ref{fig4}) that the most important factors for
a career in astronomy are networking, the number of papers, to work on
a hot topic, the quality of research, and a regular presence at
conferences.  Networking and regular presence at conferences were
favored by all female professors.

\subsubsection{Contentment with current job}

\begin{figure}[t]
\includegraphics[width=8.3cm,angle=0,clip]{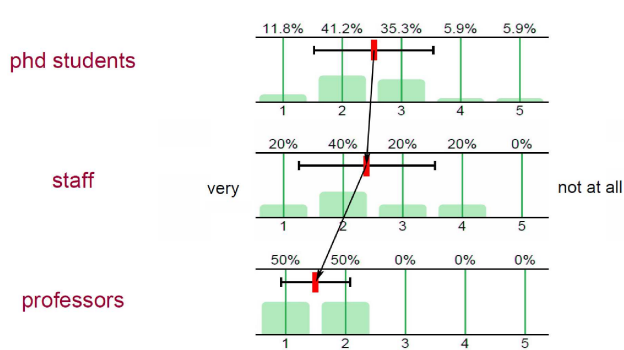}
\caption{(online colour at: www.an-journal.org) ``All in all, how content are
you with your current job situation?" Left is ``very", right ``not at all".  
Percentages and mean values (red bar) for different career levels.}
\label{fig5}
\end{figure}

\begin{figure}[t]
\includegraphics[width=8.3cm,angle=0,clip]{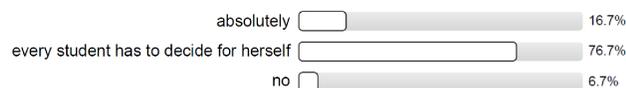}
\caption{``Would you recommend to young female students to strive for a career in astronomy?"}
\label{fig6}
\end{figure}

The contentment with the current job situation of female astronomers
is good and even shifting from okay to very good with each career
advance (Fig.~\ref{fig5}). Nonetheless only 17\,\% would recommend to young
female students to strive for a career in astronomy (Fig.~\ref{fig6}). Most
(77\,\%) say ``every student has to decide for herself", only 7\% say
``no". This seems to be in contrast to the fact that more than 90\,\%
state that they would like to continue astronomical research and their
overall contentment. It can be explained with the insecure funding
situation, long-term perspective and mobility constraints or just that
for some women both statements reflect independent issues.

\subsubsection{What was helpful for your career?}

All women in the sample were asked for their experience on what was
helpful for their career in astronomy as a recommendation to younger
women. 

Helpful is/was
\begin{itemize}
\item a motivating, encouraging, acknowledging boss/super\-visor who was a good mentor and trusted in abilities, 
and who helped getting hands on excellent data and who introduced into networks, 
\item finding projects as well as self-motivation and working independently, 
\item having role models for different topics and life phases, 
\item attending and giving talks at conferences, colloquia and seminars, 
\item successful applications for grants, observing time and soft money, 
\item stays abroad and flexibility, and 
\item colleagues who helped to advance.
\end{itemize}

\subsubsection{What was/is harmful for the career?}

We asked all participants to describe their experience on what is
harmful for a career in astronomy. The answers, although very
different for individuals, show the contrary of the above
recommendations. 

Harmful is
\begin{itemize}
\item unreliable line-managers/supervisors,
\item bad advisor/no strong mentor/no strong group (resulting in bad PhD, PhD/habilitation takes too long, 
lack of papers, no collaborations, no counseling, lack of support, no networking),
\item  change of PhD-advisor, 
\item not being able to plan freely/choose a job anywhere (due to partner or children, maternity breaks), 
\item no/not enough stays abroad (after PhD!), 
\item working too much in teaching/management/service, 
\item staying too long at same place, being old, 
\item lack of self-promotion, 
\item not working on a hot topic, 
\item working in big collaborations (depending on the authorship policy).
\end{itemize}

\subsection{A backward time-dilatation of German society?}

This part of the survey was meant to 
seeing gender
issues less dead-serious. Unfortunately, the examples that were
collected are more embarrassing for our society as a whole, above
all. After a long discussion whether these comments should be
published, the participants in the 2011 AFN meeting in Heidelberg
decided to do so. Our goal is not to shock or make fun of individuals,
but rather to show what female astronomers are still confronted
with. By publishing these remarks, we hope to increase the awareness
that comments like these are perceived as discriminating.

The examples fall in three categories:

\begin{enumerate}
\item{General designation (unconscious or conscious prejudice):}\\
{\it 
\begin{description}
 \item[1.1.] I know you would like to work, but if all women would stay at home, we would have much less unemployment.
 \item[1.2.] For a woman your seminar was good.
 \item[1.3.] You must be the secretary.
 \item[1.4.] Female scientists are more masculine than normal women.
 \item[1.5.] Special programs for women discriminate men.
 \item[1.6.] Good morning gentlemen.
 \item[1.7.] Dear Sir.
 \item[1.8.] Ha ha, that is the alibi/quota woman! \\[-2mm]
\end{description}
 } 
 
\item{Women are not treated independently of their partner:}\\
 {\it 
 \begin{description}
 \item[2.1.] The husband of this (female) applicant has a better position, so she does not need a job.
 \item[2.2.] Why you want more money? Your husband is working!
 \item[2.3.] Will you stop your PhD education now that you married?\\[-2mm]
 \end{description}
  }   
  
\item{Pressing into the mother-role:}\\
 {\it 
 \begin{description}
 \item[3.1.] You have a diploma [i.e., M.Sc. degree], why do you also want a PhD? 
  Now you can go home and have children.
  \item[3.2.] Women who give birth dont come back.
  \item[3.3.] To a woman with children:\\
     The permanent position is for mister XY, he has to support his family.
  \item[3.4.] She wouldn't come anyway (for a job) due to the children.
  \item[3.5.] It is better for the children if the mother stays at home.
\end{description}
}
\end{enumerate}

Since the above statements represent a collection of comments from our survey, we do not know 
in which context these have been verbalized. Some comments have been repeated several times and perceived as more unfair with each repetition. Even if some of those comments were not meant to be serious we conclude that role stereotypes and unconscious bias are still at work.      

\subsection{Example situations}

In the following we list three example situations from German universities and research institutes 
that have been reported to us.
 These examples illustrate, why female astronomers are less likely than men to proceed with a career in science after the PhD.
 Names and places are not given 
 in order to protect individuals who shared their experience wishing for a 
 positive social evolution of our science communities.
\begin{itemize}
\item[a)] Selection committee for a full professorship at a
Ger\-man university: This university has five full professorship{s} in astronomy, none of them
filled with a female astronomer.
  The female applicant gives her science presentation
  and is interrupted by the head of the selection committee shortly after the beginning. 
  The interrupting person turns to the audience saying:
  ``Nobody believes that anyway.'' During the following interview, the
  equal opportunity officer discredits the applicant in a similar
  manner by stating that the respective university did not have any
  gender issues. \\[-0.2cm]
  
\item[b)] Equally qualified female postdoc works at the same institution as her husband: 
He has a full contract including health insurance for not employed family members, and she was
told to content herself with a stipend. In contrast to a regular contract,
the stipend does not entitle her to retirement benefits, unemployment
compensation, or  parental allowance. Knowing
that she is bound to the place due to family issues, the employer put her
on a project position where she can perform scientific research only
in a very limited frame, this way pressing her out of track. \\[-0.2cm]

\item[c)] Female scientist with children working at a German university:  
She negotiates with her supervisor to work
partly at home to use her time more efficiently. As a result, she is pressed to accept a part-time contract. 
Male colleagues with children in the same research group spend less time than her at work, but are fully paid 
without any inquiries.
\end{itemize}

Each example reflects individual situations that might not be gender specific. 
However, these are three cases, that were brought to our attention by female astronomers only. 
We note that all above institutions have signed commitments to increase the fraction of 
female scientists at all career stages as well as gender policies that should ensure
that qualified female astronomers experience equal opportunities. 

\section{Conclusion}

In this survey we study the job situation of women in astronomy in Germany and of German 
women abroad and review indicators for their career development. Our main findings can be summarized as follows:
\begin{enumerate}
\item Networking and human support are essential. 
\item Young students should carefully choose their supervisor. 
\item Postdoctoral experience abroad is a crucial factor. 
\item Motherhood complicates things. 
\item Prejudices are abundant, and are perceived as discriminating. 
\item Never trust a (job) promise until everything is fixed on paper and signed by all parties. 
\end{enumerate}

Networking is an important part of science and every-day life. Women
 find networking in professional environments challenging
as they usually juggle more than just their jobs and are still
confronted with gender prejudice. We recommend young students to
carefully choose the group they want to work with and to regularly
attend and give talks at conferences. Since surveys (\cite{koene};
\cite{ivie}; \cite{fohlmeister}) have shown that women in general are
at risk to receive less opportunities, having a good and trustworthy mentor can be
extremely helpful for the advancement (see studys also studies by \cite{tsui}; \cite{tyson}). We find that almost every
second participant in our survey found their current job due to
personal contacts. This means that young students increase their
chances by having a good network. Astronomy institutions should put
efforts to support female potentials in this respect as history shows
that this comes naturally more often only for men. Here we do not want
to give the impression that networking is more important than an
excellent publication output, successful fund-raising and scientific
standing. But nevertheless it is very important for building
successful collaborations and social-scientific contacts that help to
reach these goals. Many measures for scientific success, as the number of citations
or invitations to give talks, also depend on networking.  

Most  universities and research institutes in Germany have equal opportunities statutes.
This includes  procedures of international standard, like involving external committees when 
hiring people on high-level positions, and committees composed of a diversity of members including women. However, 
there have been reported cases, where written equal opportunity clauses were ignored or 
   the equal opportunity representatives    (being sometimes the secretary, undergraduate 
   students or women on short-term contract)  had no power to implement them.  
   
Although experience abroad is essential for a career in astronomy,
a considerable fraction of young female astronomers seem to prefer to be close to family and
friends when choosing a job. When it comes to motherhood women get
restricted in mobility. Additionally, they are more often confronted
with prejudice.  The overall contentment with the work of women in
astronomy is good and even improves with each career advance. More
than 40\,\% of the female astronomers in the sample state that they want
to continue research in astronomy on the long term. 
Most female astronomers state that young female students have to decide 
themselves if they want to strive for a career in astronomy; only 17\,\%
would recommend it.

While this survey, by design, only covered `easily measurable' issues of career development, 
including some prejudices and conscious biases, we would like to mention, that the broad field of 
unconscious biases can be as important or harmful a factor for career development, as the topics discussed here
 (see \cite{urry} and references therein).

We conclude that in order to make it more attractive for women to stay in astronomy, 
the promotion prospects, job safety and family support should be improved. A socially evolved 
working environment including appropriate mentoring schemes should not be underestimated as examples from outside Germany show. 

No comparable data exist for male counterparts in the German astronomy society. This makes it more difficult
 to judge whether certain results are specific to female astronomers and to assess the impact of multiple causes
 such as being female {\it and} having a family. It would also be interesting to study to which extent the
 career path choices and the private life differ between female and male astronomers. 
We recommend to collect similar data for a representative sample of female and male astronomers within a 
larger framework such as the Astronomische Gesellschaft.

\acknowledgements
We would like to thank Jessica Agarwal, Eva Grebel, Inga Kamp, Stefanie Komossa, Sabine Reffert, and Sonja Schuh 
for valuable comments, Rebekka Weinel for technical support and everyone who filled the questionnaire.

\end{document}